\documentclass[conference]{IEEEtran}

\usepackage[colorlinks,urlcolor=blue,linkcolor=blue,citecolor=blue]{hyperref}
\expandafter\def\expandafter\UrlBreaks\expandafter{\UrlBreaks\do\/\do\*\do\-\do\~\do\'\do\"\do\-}
\usepackage{cite}
\usepackage{tikz}
\usepackage{graphicx}

\begin{document}

\title{The Evolution of Decentralized Systems: From Gray's Framework to Blockchain and Beyond}

\author{
\IEEEauthorblockN{Zhongli Dong}
\IEEEauthorblockA{The University of Sydney\\
Sydney, NSW 2006, Australia\\
zhongli.dong@sydney.edu.au}
\and
\IEEEauthorblockN{Young Choon Lee}
\IEEEauthorblockA{Macquarie University\\
Sydney, NSW 2109, Australia\\
young.lee@mq.edu.au}
\and
\IEEEauthorblockN{Albert Y. Zomaya}
\IEEEauthorblockA{The University of Sydney\\
Sydney, NSW 2006, Australia\\
albert.zomaya@sydney.edu.au}
}

\maketitle

\begin{abstract}
Blockchain technology is often discussed as if it emerged from nowhere, yet its architectural DNA traces directly to the decentralized computing principles James~N.\ Gray articulated in 1986. This paper maps the conceptual lineage from Gray's requestor/server model to modern blockchain architectures, showing how his emphasis on modularity, autonomy, data integrity, and standardized communication anticipated the design of systems like Bitcoin and Ethereum, and, more recently, the Web3 movement and Layer-2 scaling architectures. We examine consensus mechanisms, cryptographic foundations, rollup-based Layer-2 protocols, and cross-chain interoperability through this historical lens, identify persistent challenges in scalability and modularity, and outline future directions toward Web4: an intelligent, decentralized internet integrating blockchain, artificial intelligence, and the Internet of Things.
\end{abstract}

\begin{IEEEkeywords}
Blockchain, decentralized systems, Web3, Web4, Layer-2, consensus mechanisms, smart contracts, interoperability
\end{IEEEkeywords}

\section{Introduction}

Decentralized systems have evolved from an architectural curiosity of the 1980s into a defining paradigm of modern computing. In centralized architectures, a single server or cluster governs all data processing and decision-making, an arrangement that creates bottlenecks, single points of failure, and inherent scalability limits. The shift toward distributing control across independent, cooperating nodes promised to address each of these weaknesses, and the intervening decades have largely vindicated that promise.

A pivotal moment in this shift was James~N.\ Gray's 1986 paper on decentralized computer systems~\cite{gray1986}. Writing at Tandem Computers, Gray laid out a requestor/server architecture in which system functions were encapsulated as modular services communicating via standardized message protocols. His framework emphasized four principles that remain strikingly relevant: \emph{modularity} (independent, replaceable components), \emph{autonomy} (each node controls its own data and processes), \emph{data integrity} (consistent, tamper-resistant records across distributed copies), and \emph{standardized communication} (common protocols enabling seamless interaction among heterogeneous components).

Two decades later, Satoshi Nakamoto's Bitcoin white paper~\cite{nakamoto2008bitcoin} demonstrated that these very principles could underpin a global, trustless financial system. Blockchain, a distributed ledger maintained by a peer-to-peer network with no central authority, realizes Gray's vision in a concrete, widely deployed form, while adding innovations in cryptography, consensus, and smart-contract execution that Gray could not have anticipated. More recently, the Web3 movement has extended this trajectory to the consumer internet, and an emerging generation of Layer-2 scaling protocols has begun to address the throughput limitations that have long constrained blockchain adoption.

This paper traces the conceptual lineage from Gray's 1986 framework through blockchain and Web3 to the emerging vision of Web4. We examine how each of Gray's four principles manifests in blockchain design, highlight the technical mechanisms that bridge theory and practice, survey real-world applications, and identify the open challenges that will shape the next generation of decentralized systems.

\section{Gray's Framework: Principles That Endure}

Gray's rationale for decentralization was both organizational and technical. Large enterprises of the 1980s needed to distribute control across geographically dispersed units while ensuring that each unit retained autonomy over its local data and processes. Technically, decentralization addressed capacity limits, reduced response latency by placing processing closer to the data source, and improved fault tolerance by eliminating single points of failure~\cite{gray1986}.

His requestor/server model was deliberately modular: each server exposed high-level operations on well-defined objects, and requestors accessed these operations through a standardized message protocol. This separation of concerns meant that individual servers could be upgraded, replaced, or replicated without disrupting the rest of the system, a design principle that anticipated the microservices architectures prevalent in cloud computing today.

Gray was candid about the costs of decentralization. Coordination across autonomous nodes introduced complexity; maintaining consistency in distributed data required sophisticated algorithms; and the overall system demanded more careful design than its centralized counterpart. These trade-offs (autonomy versus coordination, performance versus consistency) have persisted through every subsequent generation of distributed systems, including blockchain.

Gray also emphasized the importance of transaction management in decentralized environments. He argued that the key to reliable distributed systems was the ability to ensure that operations either completed entirely or had no effect, that is, the atomicity property that would later become central to database theory and, eventually, to blockchain's model of state transitions. His insight that the main technical problem unique to decentralized systems is the lack of a common memory foreshadowed the consensus challenge that blockchain would later solve through cryptographic means.

The principles Gray identified have since been validated at enormous scale. The internet itself is a decentralized network with no single point of control. Peer-to-peer file-sharing systems such as BitTorrent demonstrated that decentralization could achieve global-scale data distribution. Cloud platforms distribute computing resources across worldwide data centers, offering elastic scalability that would have been inconceivable in Gray's era. In each case, the recurring themes are modularity, autonomy, integrity, and standardization, the very pillars of Gray's 1986 framework.

\section{Blockchain as Realization}

\begin{figure}[t]
\centering
\begin{tikzpicture}[scale=0.68, every node/.style={transform shape}]
    \draw[thick, fill=blue!15, rounded corners=2pt] (0,0) rectangle (3.2,2.4);
    \node[font=\bfseries] at (1.6,1.9) {Block \#1};
    \node[font=\small] at (1.6,1.2) {Hash: 0x0000};
    \node[font=\small] at (1.6,0.6) {Prev: ---};
    \draw[thick, fill=green!15, rounded corners=2pt] (4.2,0) rectangle (7.4,2.4);
    \node[font=\bfseries] at (5.8,1.9) {Block \#2};
    \node[font=\small] at (5.8,1.2) {Hash: 0x1A3F};
    \node[font=\small] at (5.8,0.6) {Prev: 0x0000};
    \draw[thick, fill=red!15, rounded corners=2pt] (8.4,0) rectangle (11.6,2.4);
    \node[font=\bfseries] at (10.0,1.9) {Block \#3};
    \node[font=\small] at (10.0,1.2) {Hash: 0x7B2E};
    \node[font=\small] at (10.0,0.6) {Prev: 0x1A3F};
    \draw[thick, ->, >=stealth] (3.2,1.2) -- (4.2,1.2);
    \draw[thick, ->, >=stealth] (7.4,1.2) -- (8.4,1.2);
\end{tikzpicture}
\caption{Basic structure of a blockchain. Each block contains a cryptographic hash of the previous block, forming a tamper-evident chain. Altering any block invalidates all subsequent hashes.}
\label{fig:blockchain}
\end{figure}
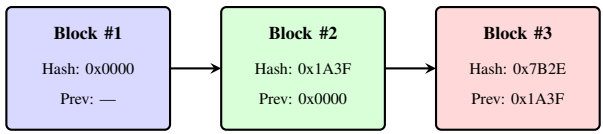

Blockchain technology translates Gray's abstract architectural principles into a concrete, globally deployed system. Introduced through the Bitcoin white paper in 2008~\cite{nakamoto2008bitcoin}, blockchain is a distributed ledger in which every participating node maintains a copy of the transaction history. New records are grouped into blocks, each cryptographically linked to its predecessor, forming a tamper-evident chain (see Fig.~\ref{fig:blockchain}). No central authority governs the ledger; instead, a consensus mechanism ensures that all honest nodes converge on the same state.

The mapping to Gray's principles is direct. Blockchain nodes are autonomous: each independently validates transactions, maintains its own copy of the ledger, and can join or leave the network at will. The system is modular: consensus layers, networking layers, and application layers can be designed, upgraded, and replaced independently, as Ethereum's transition from Proof of Work (PoW) to Proof of Stake (PoS) demonstrated~\cite{buterin2014next}. Data integrity is enforced by cryptographic hashing and Merkle trees, which make any retroactive alteration computationally infeasible. Standardized protocols, from Bitcoin's peer-to-peer gossip protocol to Ethereum's widely adopted ERC-20 token standard, ensure that diverse implementations can interoperate.

Ethereum extended the blockchain paradigm beyond simple value transfer by introducing smart contracts: self-executing programs stored on the blockchain that run when predetermined conditions are met~\cite{buterin2014next}. This programmability transformed blockchain from a specialized financial ledger into a general-purpose decentralized computing platform, enabling applications in supply-chain management, healthcare, finance, and digital identity. Smart contracts also introduced the notion of \emph{composability}: individual contracts can call one another, creating complex workflows from simple building blocks, a direct echo of Gray's modular server architecture.

\subsection{From Blockchain to Web3}

The programmability of platforms like Ethereum catalysed a broader architectural vision known as Web3: a re-imagining of the internet in which users, rather than platform operators, own their data, digital assets, and online identities. Where Web~1.0 was read-only and Web~2.0 introduced user-generated content controlled by centralized platforms, Web3 replaces platform intermediaries with blockchain-based protocols, smart contracts, and token-economic incentive layers~\cite{wang2024web3}.

In a Web3 architecture, decentralized applications (dApps) interact with on-chain smart contracts instead of proprietary back-end servers. User authentication relies on cryptographic wallets rather than platform-managed credentials. Data can be stored on decentralized networks such as IPFS or Filecoin rather than in corporate data centers. Each of these design choices maps directly to Gray's principles: modularity (composable smart contracts replace monolithic platforms), autonomy (users hold their own keys and data), integrity (on-chain state is tamper-evident), and standardized communication (token standards and open application binary interfaces enable permissionless interoperability among dApps).

Web3 remains a work in progress. Usability barriers, regulatory uncertainty, and the scalability constraints of underlying blockchains have slowed mainstream adoption. The user experience of managing private keys, paying gas fees, and navigating wallet interfaces remains far more complex than the password-based authentication of Web~2.0 platforms. Nonetheless, the architectural direction is clear: the same decentralization principles that Gray advocated for enterprise computing are now being applied to reshape the consumer internet itself.

\section{Technical Bridges: Consensus, Cryptography, and Communication}

Three technical pillars bridge Gray's conceptual framework and modern blockchain implementations.

\subsection{Consensus Mechanisms}

Achieving agreement among autonomous, potentially adversarial nodes is blockchain's central challenge, and the modern incarnation of the coordination problem Gray identified in 1986. Bitcoin's PoW mechanism requires miners to solve computationally expensive cryptographic puzzles, ensuring that appending a block demands significant effort and making attacks economically prohibitive~\cite{nakamoto2008bitcoin}. The difficulty of PoW puzzles adjusts dynamically based on the network's total computational power, maintaining a consistent block interval regardless of how many miners participate.

While effective, PoW's energy consumption has prompted the development of alternatives. PoS selects validators based on the cryptocurrency they commit as collateral, dramatically reducing energy requirements while preserving security guarantees through economic incentives: validators who approve fraudulent transactions lose their staked assets. Ethereum's ``Merge'' in September 2022 demonstrated that a large-scale blockchain could transition from PoW to PoS without disrupting its ecosystem, reducing the network's energy consumption by an estimated 99.95\%.

More recent designs such as Algorand's Byzantine Agreement protocol achieve high throughput without sacrificing decentralization by using verifiable random functions to select block proposers and committee members~\cite{gilad2017algorand}. Delegated Proof of Stake systems trade some decentralization for performance by concentrating validation authority in a smaller set of elected delegates. Each mechanism represents a different trade-off among decentralization, security, and performance, the same tension Gray identified in 1986.

\subsection{Cryptographic Foundations}

Data integrity in blockchain relies on two cryptographic primitives. First, every block contains a hash of its predecessor, creating a chain in which any modification propagates detectable changes through all subsequent blocks. Cryptographic hash functions (such as SHA-256 in Bitcoin) produce a fixed-size digest from arbitrary input data; even a single-bit change in the input produces a completely different hash, making tampering immediately detectable.

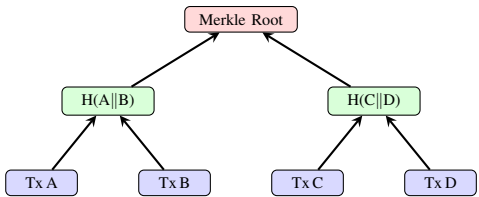
\begin{figure}[t]
\centering
\begin{tikzpicture}[scale=0.68, every node/.style={transform shape}]
    \node (L1) at (0,0) [draw, rectangle, fill=blue!15, rounded corners=2pt, minimum width=1.4cm, minimum height=0.5cm] {\small Tx\,A};
    \node (L2) at (2.6,0) [draw, rectangle, fill=blue!15, rounded corners=2pt, minimum width=1.4cm, minimum height=0.5cm] {\small Tx\,B};
    \node (L3) at (5.2,0) [draw, rectangle, fill=blue!15, rounded corners=2pt, minimum width=1.4cm, minimum height=0.5cm] {\small Tx\,C};
    \node (L4) at (7.8,0) [draw, rectangle, fill=blue!15, rounded corners=2pt, minimum width=1.4cm, minimum height=0.5cm] {\small Tx\,D};
    \node (M1) at (1.3,1.6) [draw, rectangle, fill=green!15, rounded corners=2pt, minimum width=1.8cm, minimum height=0.5cm] {\small H(A$\|$B)};
    \node (M2) at (6.5,1.6) [draw, rectangle, fill=green!15, rounded corners=2pt, minimum width=1.8cm, minimum height=0.5cm] {\small H(C$\|$D)};
    \node (R) at (3.9,3.2) [draw, rectangle, fill=red!15, rounded corners=2pt, minimum width=2.2cm, minimum height=0.5cm] {\small Merkle Root};
    \draw[thick, ->, >=stealth] (L1) -- (M1);
    \draw[thick, ->, >=stealth] (L2) -- (M1);
    \draw[thick, ->, >=stealth] (L3) -- (M2);
    \draw[thick, ->, >=stealth] (L4) -- (M2);
    \draw[thick, ->, >=stealth] (M1) -- (R);
    \draw[thick, ->, >=stealth] (M2) -- (R);
\end{tikzpicture}
\caption{Merkle tree. Transactions are hashed pairwise until a single root hash is produced. Verifying any transaction requires only a logarithmic number of hash checks.}
\label{fig:merkle}
\end{figure}

To efficiently verify the integrity of large sets of transactions within a single block, blockchains employ Merkle trees (see Fig.~\ref{fig:merkle}): a binary tree structure in which each leaf node is a hash of a transaction, and each non-leaf node is the hash of its two children. This hierarchical structure allows any individual transaction to be verified by checking only a logarithmic number of hashes rather than re-hashing the entire block.

Second, the Elliptic Curve Digital Signature Algorithm (ECDSA) enables users to sign transactions with a private key and have them verified by the network using the corresponding public key. When a user initiates a transaction, they produce a digital signature over the transaction data. Network nodes verify this signature using the sender's public key, confirming both authenticity and integrity. The security of ECDSA rests on the computational infeasibility of the Elliptic Curve Discrete Logarithm Problem: given a public key, deriving the corresponding private key is believed to require an astronomically large number of operations with current technology~\cite{koblitz1987elliptic}. Together, hashing and digital signatures provide the tamper-evidence and authentication that Gray's framework required but lacked a concrete mechanism to enforce at scale.

\subsection{Layer-2 Scaling Architectures}

Gray observed that decentralized systems must accommodate growth without requiring a complete redesign. In blockchain, this concern has given rise to Layer-2 (L2) solutions: protocols built on top of a base-layer (Layer-1) blockchain that offload transaction execution while inheriting the security guarantees of the underlying chain~\cite{thibault2022rollups}.

Three families of L2 architectures have emerged. \emph{State channels} (including Bitcoin's Lightning Network~\cite{poon2016bitcoin}) allow two parties to conduct an unlimited number of off-chain transactions, settling only the net result on Layer~1. This approach is highly efficient for repeated bilateral interactions but does not generalize easily to complex multi-party computations.

\emph{Sidechains} operate as independent blockchains with their own consensus mechanisms but periodically anchor state commitments to the parent chain. They offer flexibility, allowing developers to experiment with different virtual machines, consensus algorithms, and block parameters, at the cost of weaker security guarantees, since sidechain validators are typically a smaller, separate set from Layer-1 validators.

\emph{Rollups}, the dominant L2 paradigm today, execute transactions off-chain, compress them into batches, and post the batch data back to Layer~1. Optimistic rollups assume batch validity and allow a challenge period during which any observer can submit a fraud proof. Zero-knowledge (ZK) rollups instead attach a succinct cryptographic validity proof that Layer~1 can verify in constant time, enabling immediate finality without a challenge window~\cite{thibault2022rollups}. ZK rollups offer stronger security guarantees and faster withdrawals, but generating validity proofs is computationally expensive and introduces prover latency.

The modular separation between execution (L2) and settlement/data availability (L1) is a direct instantiation of Gray's modularity principle: each layer can be optimized, upgraded, or replaced independently. Ethereum's roadmap explicitly embraces this modular thesis, splitting responsibilities across a consensus layer, an execution layer, and a forthcoming data-availability layer (danksharding). This layered architecture echoes Gray's requestor/server model at internet scale.

\subsection{Interoperability and Cross-Chain Communication}

Gray stressed that standardized communication was essential for heterogeneous components to cooperate. In blockchain, this concern manifests as the interoperability problem: how can independent blockchain networks exchange assets and data without relying on a trusted intermediary?

Several approaches have emerged. Atomic swaps use hashed time-locked contracts to enable trustless token exchange between chains~\cite{herlihy2018atomic}. Relay-chain architectures such as Polkadot connect multiple specialized blockchains (parachains) through a shared security model~\cite{wood2016polkadot}. Hub-and-spoke models such as Cosmos employ an Inter-Blockchain Communication protocol to link sovereign chains, each retaining its own consensus and governance. A comprehensive survey catalogs these and other approaches, identifying open problems in security assumptions, latency, and composability~\cite{belchior2021survey}.

The proliferation of interoperability solutions underscores Gray's insight that standardization is not merely a convenience but an architectural necessity. Without common protocols for cross-chain messaging, the blockchain ecosystem risks fragmenting into isolated silos, the very outcome that decentralization was meant to prevent.

\section{Applications and Impact}

Blockchain's combination of autonomy, integrity, and transparency has found traction across multiple industries. In each case, the value proposition traces back to Gray's original argument: removing central bottlenecks, distributing control, and ensuring data integrity across independent participants.

\textbf{Supply-chain management.} Walmart's collaboration with IBM on the Food Trust platform demonstrated that blockchain could reduce the time to trace a food product's origin from seven days to 2.2 seconds~\cite{kamath2018food}. By recording every step of a product's journey on an immutable ledger, supply-chain participants gain end-to-end visibility, improving safety, reducing fraud, and ensuring regulatory compliance. The modular design of the platform allows new partners to join without altering existing infrastructure, serving as a practical demonstration of Gray's modularity principle.

\textbf{Healthcare.} Projects such as MedRec use blockchain to create patient-controlled electronic health records, enabling secure data sharing among providers while preventing unauthorized access and data breaches~\cite{ekblaw2016case}. The decentralized architecture ensures that no single institution controls the patient's data, and cryptographic access controls allow patients to grant and revoke access at a granular level.

\textbf{Decentralized finance (DeFi).} Smart-contract platforms have given rise to an ecosystem of lending, borrowing, and trading services that operate without traditional financial intermediaries, lowering costs and broadening access~\cite{schar2021decentralized}. DeFi protocols such as Aave and Uniswap are composable: they can be combined to create complex financial products, a property that directly mirrors the modular composability Gray advocated. As of 2024, tens of billions of dollars in digital assets are managed by DeFi protocols, demonstrating that decentralized systems can operate at meaningful financial scale.

\textbf{Digital identity.} Self-sovereign identity systems built on blockchain allow individuals to control their own credentials and share them selectively with verifiers, reducing the risk of centralized data breaches. Unlike traditional identity providers, these architectures ensure that no single party holds a complete picture of a user's identity; instead, cryptographic proofs allow users to attest to specific attributes without revealing unnecessary personal information.

\section{Open Challenges}

Despite significant progress, several challenges limit the broader adoption of blockchain technology.

\textbf{Scalability.} Despite the Layer-2 architectures described above, public blockchains still face a fundamental throughput gap relative to centralized payment networks. Bitcoin processes approximately 7 transactions per second; Ethereum's Layer-1 handles around 30. By contrast, Visa processes more than 24,000 transactions per second. Sharding (partitioning the network into parallel segments) remains a promising but technically challenging Layer-1 approach~\cite{zamani2018rapidchain}. ZK-rollups continue to improve, but prover latency, sequencer centralization, and cross-rollup composability are unresolved issues. Achieving scalability without sacrificing decentralization or security, commonly referred to as the blockchain trilemma, remains the field's defining open problem.

\textbf{Modularity at scale.} While Ethereum's layered architecture exemplifies modularity, achieving clean separation of concerns across consensus, execution, and data-availability layers remains an active research area. The interfaces between layers must be precisely defined and rigorously secured; a vulnerability at any inter-layer boundary can compromise the entire stack.

\textbf{Energy and sustainability.} PoW's energy consumption has drawn justified criticism. Ethereum's transition to PoS dramatically reduced its energy footprint, but PoW remains dominant for Bitcoin and several other major chains. The long-term security properties of PoS under adversarial conditions require continued scrutiny.

\textbf{Governance and regulation.} Decentralized systems challenge existing regulatory frameworks. Establishing accountability in systems without a central operator, ensuring compliance with data-protection laws such as the GDPR's ``right to erasure'' on an immutable ledger, and resolving disputes in decentralized governance structures remain largely unsolved.

\textbf{Usability.} The gap between blockchain's technical potential and its real-world usability remains wide. Managing cryptographic keys, understanding gas fees, and navigating the fragmented landscape of chains, wallets, and bridges are significant barriers for non-technical users. Until the user experience approaches the simplicity of centralized alternatives, mass adoption will remain elusive.

\section{Future Directions: Toward Web4}

The next phase of decentralized systems will likely be defined by convergence, not just of blockchain with individual technologies, but of an entire stack that fuses decentralized infrastructure, artificial intelligence (AI), spatial computing, and ubiquitous connectivity into what researchers are beginning to call Web4~\cite{wang2024web3}.

\textbf{From Web3 to Web4.} If Web3 decentralizes ownership and trust, Web4 aims to make that decentralized substrate intelligent and autonomous. In a Web4 architecture, AI agents, and not just human users, become first-class participants in blockchain networks: negotiating smart contracts, managing decentralized identities, and orchestrating resources across chains. The vision combines the user-sovereignty of Web3 with the adaptive intelligence of large-scale AI, the real-time sensing of IoT, and the immersive interfaces of spatial computing. Gray's four principles provide the architectural scaffolding: modularity enables the independent evolution of AI, blockchain, and IoT layers; autonomy supports self-governing AI agents operating without centralized oversight; integrity ensures verifiable provenance for AI training data and model outputs; and standardized communication enables seamless interoperability among heterogeneous agents, devices, and chains.

\textbf{AI and blockchain synergy.} AI models increasingly rely on large, distributed datasets whose provenance and integrity are difficult to verify in centralized systems. Blockchain can provide tamper-evident training-data registries, transparent model-audit trails, and decentralized marketplaces for data and compute. Conversely, AI techniques can optimise blockchain parameters such as gas pricing, sequencer ordering, and cross-shard routing in real time, creating a feedback loop between intelligence and infrastructure. The emergence of on-chain AI agents that autonomously execute trading strategies and manage liquidity illustrates how this convergence is already materializing.

\textbf{Internet of Things (IoT).} IoT devices generate vast, distributed data streams naturally suited to decentralized management. Blockchain-anchored IoT networks can provide automated device-to-device transactions via smart contracts, decentralized device identity, and tamper-evident sensor logging. The challenge lies in adapting blockchain and Layer-2 protocols to the resource constraints and low-latency requirements of edge environments. Lightweight consensus mechanisms and purpose-built L2 chains for IoT telemetry represent active areas of research.

\textbf{Layer-2 and Layer-3 evolution.} The modular blockchain thesis is likely to intensify, with specialized Layer-2 and Layer-3 chains optimized for specific workloads (high-frequency DeFi, AI inference, IoT telemetry, gaming) while sharing a common settlement and data-availability layer. Decentralized sequencing, shared proving infrastructure, and cross-rollup messaging protocols are active research frontiers. The emergence of ``rollup-as-a-service'' platforms suggests that the barrier to creating application-specific chains will continue to fall.

\textbf{Standardization.} As the Web4 stack matures, universally accepted standards for data formats, communication protocols, smart-contract interfaces, and AI-agent interaction will become essential for interoperability and regulatory compliance.

Gray's four principles (modularity, autonomy, integrity, and standardized communication) will remain the architectural backbone of these future systems, even as the specific technologies evolve.

\section{Conclusion}

The architectural principles James~N.\ Gray articulated in 1986 have proven remarkably durable. Blockchain technology did not emerge in a vacuum; it is the concrete realization of a vision for decentralized computing that prioritizes modular design, node autonomy, data integrity, and standardized communication. The Web3 movement extends this vision to the consumer internet, Layer-2 architectures instantiate Gray's modularity at protocol scale, and the emerging Web4 paradigm promises to fuse decentralized infrastructure with artificial intelligence and ubiquitous connectivity. Scalability, interoperability, governance, and usability remain open problems, but the foundational principles that guided Gray's work continue to provide a reliable compass for researchers and practitioners building the next generation of decentralized systems.

\section*{Acknowledgment}
This work was supported by the Australian Research Council (ARC) Research Hub for Future Digital Manufacturing (IH230100013).

\bibliographystyle{IEEEtran}
\bibliography{references}

\end{document}